\documentclass[twocolumn,aps,showpacs,nofootinbib,epsfig]{revtex4}
\usepackage{amssymb}
\usepackage{amsmath}
\usepackage{graphics}
\usepackage{epsfig}
\usepackage[dvips]{color}

\newcommand{\be}{\begin{equation}}
\newcommand{\ee}{\end{equation}}
\newcommand{\bea}{\begin{eqnarray}}
\newcommand{\eea}{\end{eqnarray}}
\newcommand{\beas}{\begin{eqnarray*}}
\newcommand{\eeas}{\end{eqnarray*}}

\begin{document}
\title{The QCD phase diagram from Schwinger-Dyson Equations}
\author{Enif
  Guti\'errez$^1$, Aftab Ahmad$^{1,2}$, Alejandro Ayala$^3$, Adnan Bashir$^1$, and Alfredo Raya$^1$}
\affiliation{$^1$Instituto de F\1sica y Matem\'aticas, Universidad
Michoacana de San Nicol\'as de Hidalgo, Edificio C-3, Ciudad
Universitaria, Morelia, Michoac\'an 58040, M\'exico. \\
$^2$Department of Physics, Gomal University~29220, D.~I.~Khan,
K.P.K., Pakistan. \\
$^3$Instituto de Ciencias Nucleares, Universidad Nacional
Aut\'onoma de M\'exico, Apartado Postal 70-543, M\'exico Distrito
Federal 04510, Mexico.}
\begin{abstract}

We study the phase diagram of quantum chromodynamics (QCD). For this purpose we employ the
Schwinger-Dyson equations (SDEs) technique and construct a truncation of
the infinite tower of equations by demanding a matching with the lattice
results for the quark-anti-quark condensate at finite temperature ($T$),
for zero quark chemical potential ($\mu$), that is, the region where lattice calculations are
expected to provide reliable results. We compute
the evolution of the phase diagram away from $T=0$ for increasing values of
the chemical potential by following the evolution of the heat capacity as a function of
$T$ and $\mu$. The behavior of this thermodynamic variable clearly demonstrates the 
existence of a cross-over for $\mu$ less than a critical value. However, the heat capacity develops a singularity near $\mu \approx 0.22$ GeV marking the onslaught of  a first
order phase transition characterized by the existence of a critical point.
The critical line continues until $\mu \approx 0.53$ GeV where
$T_c=0$ and thus chiral symmetry is finally restored.

\end{abstract}


\maketitle

\section{Introduction}\label{I}

The strong interaction sector of the standard model of particle
physics involves a phase transition which is relevant to the
evolution of the early universe. At low temperatures, the
observable degrees of freedom of quantum chromodynamics (QCD) are
color-singlet hadrons whereas at high energies or temperatures the
interaction gets increasingly screened and hence weak, causing
hadrons to break up into a new phase where the dominant degrees of
freedom are the defining ingredients of perturbative QCD, namely,
quarks and gluons. This is dubbed as {\em
confinement-deconfinement phase transition}. The same physical
picture prevails when chemical potential is increased. Experiments
as well as lattice QCD computations for physical quark masses
suggest that the temperature driven transition at zero chemical
potential $\mu$ is not a thermodynamic singularity. Rather, it is
a rapid but smooth crossover from the regime describable as a
gas of hadrons, to the one characterized by quarks and
gluons~\cite{cross-over}. The $\mu$ driven transition at zero
temperature $T$ is qualitatively different. This regime cannot be
accessed through lattice studies because the computations are
hampered by the notorious sign problem. Nevertheless, a number of
different model approaches indicate that the transition in this
region is strongly first order~\cite{first-order}.

Chiral symmetry restoration is another phase transition which is
expected to occur when temperature and (or) chemical potential are
sufficiently large. This happens when the dynamically generated
component of quark masses vanishes. The pattern of chiral symmetry
breaking at zero $T$ and $\mu$ already provides evidence of the
fact that large effective quark masses within light hadrons owe
themselves primarily to the strength of the QCD interaction and
dictate both their static and dynamic properties, see for example
a recent review~\cite{SDE-Review}. Its experimental implications
for the elastic and transition form factors of mesons and baryons
form an integral part of the planned program at the 12 GeV upgrade
of the Thomas Jefferson National Accelerator Facility in
Virginia~\cite{JLab12GeV}. As the strength of the QCD interaction
diminishes with
increasing $T$ and $\mu$, only the bare quark masses survive when
these parameters exceed a critical set of values. This is referred
to as the {\em chiral symmetry breaking-chiral symmetry
restoration phase transition}.

In connection with these phase transitions, there are key
questions which, to a considerable extent, provide the motivation
for the heavy ion collision program at the Relativistic Heavy Ion
Collider (RHIC) at Brookhaven National Laboratory (BNL), the Large
Hadron Collider (LHC) at the European Centre for Nuclear Research
(CERN) and the Compressed Baryonic Matter (CBM) experiment of the
future Facility for Antiproton and Ion Research (FAIR) at
Darmstadt. These can be summarized as follows~:
 \begin{enumerate}

 \item Since the first order line originating at $T=0$ in the
 QCD phase diagram cannot end at the $\mu=0$ axis
 which corresponds to the starting point of the cross-over line,
 it must terminate somewhere in the midst of the phase diagram.
 This point is generally referred to as the critical end point. On
 one hand, it rests with the experiments to find observational
 evidence for it through extracting chemical and thermal freeze out
 temperatures, and on the other, the value of a sound theoretical
 prediction for its existence and location on the QCD phase
 diagram can hardly be exaggerated.

 \item Are the two phase transitions, corresponding to
 deconfinement and chiral symmetry restoration, coincidental?
 As they both owe themselves to the diminishing of the QCD
 interaction strength, one may expect them to chart out the
 same curve in the QCD phase diagram. However, its prediction and
 measurement continue to be essential as well as a challenging
 endeavor.

 \end{enumerate}

In this article, we concentrate on the exploration of the first
point.
Monte Carlo methods of lattice computations are severely
handicapped off the $\mu=0$ axis, because the
fermion determinant becomes complex and thus standard
Monte Carlo methods fail, as the integrand is no longer real
and positive definite. However, these techniques can still
be adapted to extract some information on the QCD phase diagram for
$\mu\neq0$~\cite{Lattice}. On the other hand,
the fundamental field
theoretical equations of QCD, namely, Schwinger-Dyson Equations
(SDEs), have no such restriction. As their functional derivation
makes no appeal to the smallness of the interaction strength, they
provide an ideal, natural and unified framework to explore all the
nooks and corners of the QCD phase diagram with ease. Moreover,
the bare quark masses can be set as small as required and even
studying the chirally symmetric Lagrangian is far from being
herculean~\cite{SDE}.

We study QCD phase diagram using the tools of
SDEs  at finite temperature and chemical potential. We
work with two-flavor QCD with physical up and down current quark
masses assuming isospin symmetry. The simplest two-point quark
propagator is a basic object to analyze dynamical chiral symmetry
breaking and confinement. At finite $T$ and $\mu$, we start from
the general form of the quark propagator
 \bea
 S^{-1}(\vec{p}, \tilde{w}_n) &=& i A(\vec{p}^{\,2},\tilde{w}_n^2)
 \vec{\gamma}\cdot \vec{ p} + i \gamma_4 \tilde{w}_n^2
 C(\vec{p}^{\,2}, \tilde{w}_n^2) \nonumber \\
 &+& B (\vec{p}^{\,2}, \tilde{w}_n^2) \;,\label{ecprop2}
 \eea
 where $\widetilde{w}_n= w_n + i\mu$ and $w_n= 2(n+1)\pi T$ are the Matsubara
 frequencies. $A(\vec{p}^{\,2}, \tilde{w}_n^2)$, $B (\vec{p}^{\,2},
 \tilde{w}_n^2)$ and  $C(\vec{p}^{\,2}, \tilde{w}_n^2)$ are the scalar functions
 to be self consistently determined through solving the corresponding SDE
 \bea
 S^{-1}(\vec{p}, \tilde{w}_n)  &=& i\vec{\gamma}\cdot\vec{ p}+ i \gamma_4
 \tilde{w}_n + \Sigma(\vec{p}, \tilde{w}_n) \;, \label{ecprop}
 \eea
 where
 $\Sigma(\vec{p}, \tilde{w_n})$ is the self-energy
 expressed in terms of the dressed gluon propagator
 $D_{\mu\nu}(\vec{p}-\vec{q},\Omega_{nl})$, with
 $\Omega_{nl} = w_n - w_l$, and the full quark-gluon vertex
 $\Gamma_\mu(\vec{q},\tilde{w_l}, \vec{p},\tilde{w}_n)$ as follows
 \bea
 \Sigma(\vec{p}, \tilde{w}_n)&=&
 T\sum_{l=-\infty}^{l=\infty}  \int \frac{d^3 q}{(2\pi)^3}g^2
 D_{\mu\nu}(\vec{p}-\vec{q},\Omega_{nl}) \nonumber \\
 &\times &\frac{\lambda^a}{2}\gamma_\mu
 S(\vec{q},\tilde{w}_l)
 \frac{\lambda^a}{2}\Gamma_\nu(\vec{q},\tilde{w}_l,
 \vec{p},\tilde{w}_n)\; .
 \eea
 Here $g^2$ is the QCD interaction strength and the gluon propagator
 has the following general form in the Landau gauge
 \bea
    g^2 D_{\mu\nu}(\vec{k},\Omega_{nl})=P^T_{\mu\nu}D_T(\vec{k}^{\,2},\Omega_{nl})+P^L_{\mu\nu}D_L(\vec{k}^{\,2},\Omega_{nl})\;,
 \eea
$P^{T,L}_{\mu\nu}$ being the transversal and longitudinal
projectors defined as
\bea
 P^L_{\mu\nu}&=& \delta_{\mu\nu} -\frac{k_\mu k_\nu}{k^2} - P^T_{\mu \nu} \;, \nonumber \\
 P^T_{44}&=& P^T_{4i} = 0 ,\; \; \; \; \; P^T_{i j} =
 \delta_{i j} - \frac{k_i k_j}{\vec{k}^2} \;.
 \eea
 Following the lead of Qin {\em et. al.}~\cite{Craig}, we define
 \bea
 && \hspace{-8mm} D_{T}(\vec{k}^{\,2}, \Omega_{nl}^{2}) = {\cal D}(T) \frac{4\pi^2}{\sigma^6}(\vec{k}^{\,2}+\Omega_{nl}^{2})e^{-(\vec{k}^{\,2}+\Omega_{nl}^{2})/\sigma^2}\, ,\nonumber\\
 && \hspace{-8mm} D_{L}(\vec{k}^{\,2}, \Omega_{nl}^{2}) = {\cal D}(T) \frac{4\pi^2}{\sigma^6}(\vec{k}^{\,2}+\Omega_{nl}^{2})e^{-(\vec{k}^{\,2}+\Omega_{nl}^{2}+m_g^2)/\sigma^2} \label{dtdl}
  \eea
  where $\sigma=0.5$ GeV. There are a couple of noticeable
  differences in our adaptation of the above {\em ansatz} as will
  be clear along the way.
  Note that we take  $D_L \neq D_T $ which is generally true when $T,\mu \neq 0$.
  The difference is governed by the gluon Debye mass
  $m_g$ which enters only the longitudinal part of the gluon propagator.
  We take its value to be the lowest order hard thermal loop
  result~$ m_g^2=4 g^2 T^2/3$~\cite{Braaten:1994} with $g=1$.
  The effect of the Debye mass is
  negligible all the way through the thermal evolution, even when we use larger values of $g$. 
  The role of the Debye mass
  is to help for the convergence of the solutions.
  We also allow ${\cal D}$ to be a function of temperature.
 With appropriate traces and simplifications, we obtain
 three coupled integral equations for $A(\vec{p}^{\,2},\widetilde{w}^2_n)$,
 $B(\vec{p}^{\,2},\widetilde{w}^2_n)$ and
 $C(\vec{p}^{\,2},\widetilde{w}^2_n)$
 which are solved numerically. Convergence tests for the results
 allow us to sum over 14 Matsubara frequencies without sacrificing
 the accuracy of the results by more than 5\%. Once we have
 calculated the quark propagator, we can evaluate the quark
 anti-quark condensate as~:
 \bea
 \langle\bar{\psi} \psi \rangle = N_c \, T \sum_n \int \frac{d^3q}{(2\pi)^3}
 {\rm Tr} [S(\vec{q},\widetilde{w}_n)] \;.
 \eea
 However, notice that we have not yet said anything explicitly about ${\cal D}(T)$
 (coming from the gluon propagator) or the quark-gluon vertex, or in fact
 their product which gets projected onto the quark self-energy. Where do we extract
 this information from? At this point, we introduce the most important input ingredient
 for our calculations. As we expect lattice computations to provide reliable
 results along the $\mu=0$-axis of the QCD phase diagram,
 we make the following ansatz~:
 \bea
  {\cal D}(T) \; \Gamma_\nu(\vec{q},\tilde{w}_l,\vec{p},\tilde{w}_n)
  = D(T) \; \gamma_{\nu}  \label{V-ansatz}
 \eea
 and deduce $D(T)$ so as to reproduce the lattice results for the
 temperature dependent quark anti-quark condensate quoted
 in~\cite{latticedata}.
\begin{figure}[t!] 
\resizebox{0.52\textwidth}{!}{
\includegraphics[angle=270]{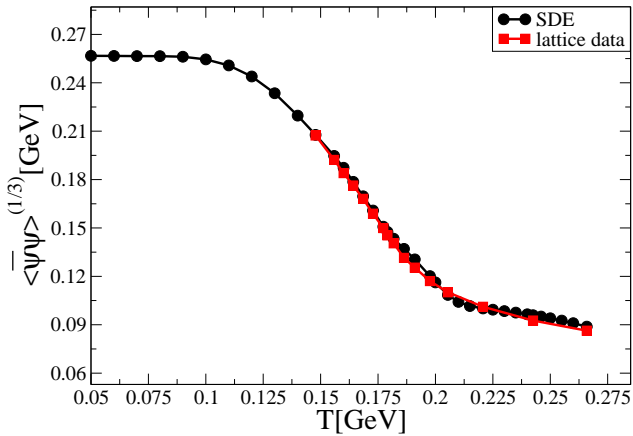}}
\caption{Function $D(T)=a-bT^3-c\tanh(d-eT^3)$ of
Eq.~(\ref{V-ansatz}) at $\mu=0$ reproduces the lattice
data~\cite{latticedata} for the quark anti-quark condensate $
\langle\bar{\psi} \psi \rangle$ very well, especially in the
region of the phase transition. \label{fig1}}
\end{figure}
Note that for the chiral limit, we reproduce the results presented
in~\cite{Craig} with appropriate modifications.
In order to introduce a finite bare mass for the quarks, we use
the fact that the mass function can be written as
$M(\vec{p}^{\,2},\widetilde{w}^2_n)=B(\vec{p}^{\,2},\widetilde{w}^2_n)/A(\vec{p}^{\,2},\widetilde{w}^2_n)$
and then make the replacement $ B(\vec{p}^{\,2},\widetilde{w}^2_n)
\longrightarrow B(\vec{p}^{\,2},\widetilde{w}^2_n) +
A(\vec{p}^{\,2},\widetilde{w}^2_n) m$. The self consistent
solution of the coupled equations subsequently yields the quark
propagator. With a fairly simple form of the function $D(T)$,
i.e.,
 \bea
  D(T)=a-bT^3-c\tanh(d-eT^3) \;,  \label{GV-fit}
 \eea
at $\mu=0$ ($a=2.17,b=343.64,c=1.76,d=0.78$ and $e=273.8$ with appropriate mass
dimensions in GeV), lattice data
is sufficiently well reproduced as shown in Fig.~\ref{fig1}.
One can draw the corresponding heat capacity curve given by
$-\partial_T \langle \bar{\psi} \psi \rangle $ which yields
a critical temperature of~$T_c \approx 0.153\,\rm GeV$, see~Fig.~\ref{fig2}. (compare it
with ~$T_c = 0.154\,\rm GeV$ reported in~\cite{latticedata}).
\begin{figure}[t!] 
\resizebox{0.52\textwidth}{!}{
\includegraphics[angle=270]{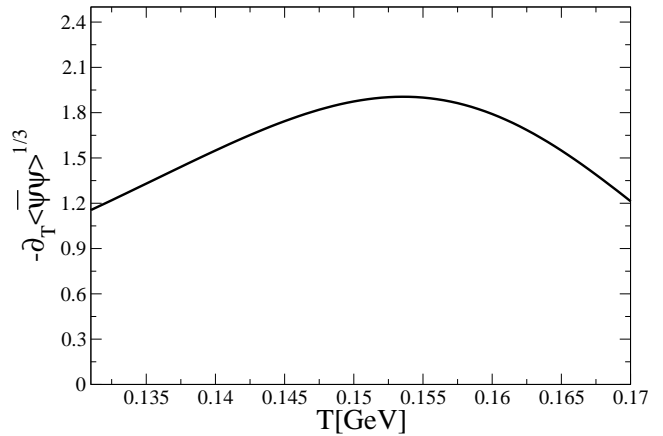}}
\caption{Heat capacity -$\partial_T \langle \bar{\psi} \psi \rangle$ calculated from
the SDE for $D(T)$ given by Eq.~(\ref{GV-fit}) as a function of temperature.
Its maximum at $T_c \sim 0.153$ GeV agrees well with~\cite{latticedata}).
\label{fig2}}
\end{figure}

\begin{figure}[t!] 
\resizebox{0.52\textwidth}{!}{
\includegraphics[angle=270]{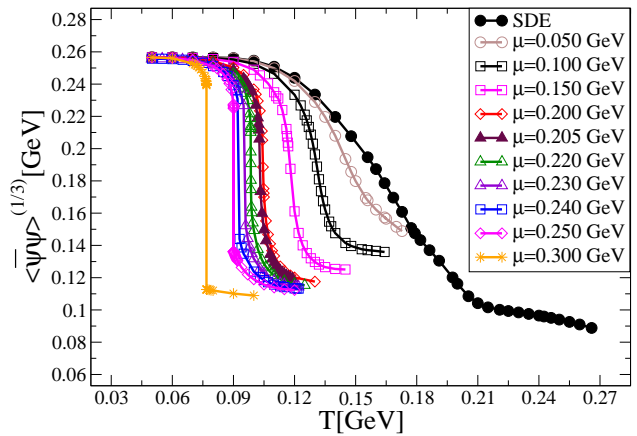}} \caption{The
condensate for different values of $\mu$ as a function
of temperature. For large $\mu$, the curve develops
a discontinuity which becomes more and more marked for
increasing values of $\mu$.
\label{fig3}}
\end{figure}

\begin{figure}[t!] 
\resizebox{0.52\textwidth}{!}{
\includegraphics[angle=270]{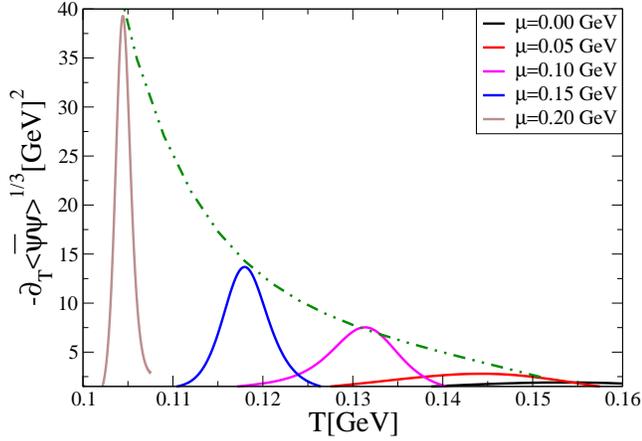}}
\caption{We plot the heat capacity curve for the condensate as a function
of temperature for different values of $\mu$. The peak gives the
critical point $(\mu_c,T_c)$. Note that the height of this thermodynamic
variable shoots up to {\em infinity} for a sufficiently large $\mu$, indicating
a change in the order of phase transition.
 \label{fig4}}
\end{figure}
We now extend our model and employ the same $D(T)$ as we evolve our
results for $\mu\neq 0$. The gluon
Debye mass becomes $\mu$-dependent, i.e., $ m_g^2=({4}T^2/3+{\mu^2}/{\pi^2})g^2$, also with $g=1$.
For increasing $\mu$, we naturally observe chiral symmetry restoration at
lower temperatures. This is depicted in Fig.~\ref{fig3} which confirms
that the effect of
incorporating the chemical potential $\mu$ tends to restore the chiral
symmetry for lower temperatures.
\begin{figure}[b!] 
\resizebox{0.52\textwidth}{!}{
\includegraphics[angle=270]{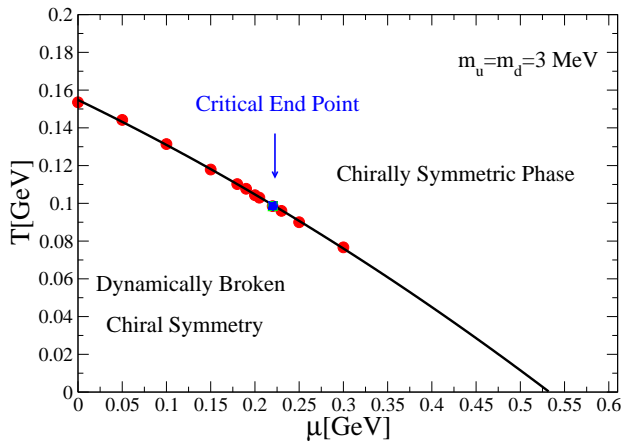}} \caption{We plot the phase diagram of QCD,
indicating also the critical end point which corresponds to $\mu \approx 0.22 GeV$
and $T=0.097 GeV$. Due to numerical difficulties above around $\mu \approx 0.3 GeV$,
we adopted a quadratic numerical fit $f_1+f_2T+f_3T^2$, to extrapolate our results to the
$T=0$ axis of the phase diagram which yields $\mu\approx 0.53\,\rm GeV$
above which chiral symmetry is restored.}
\label{fig5}
\end{figure}
An important point
to note is that as $\mu$ gets larger, a discontinuity starts to set in
for the region where the critical temperature is located. It becomes increasingly
pronounced with a relatively small variation of $\mu$. Moreover, the
slope of the curve increases drastically, so much that we eventually observe
a vertical drop, as is evident in Fig.~\ref{fig3}.
This is a likely indicator of a physical effect. In order to quantify
this change in the behavior of the heat capacity, we draw it as a function of
temperature for different values of $\mu$ to locate the critical point.
The result is shown in Fig.~\ref{fig4}. The plots become narrower for
increasing $\mu$ which helps locate the critical temperature with
progressive precision. On the other hand, the height of the curve soon
shoots up to infinity as evident from Fig.~\ref{fig4} before the chemical potential
reaches a value of about 0.22 GeV. We identify this thermodynamic singularity
with the onslaught of a first order phase transition.
Before that point is reached, the phase transition is a cross-over.
\begin{figure}[t!] 
\resizebox{0.52\textwidth}{!}{
\includegraphics[angle=270]{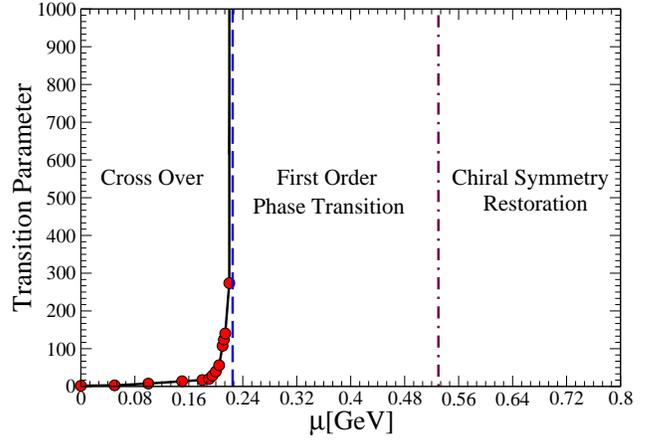}}
\caption{This plot shows different regions of the QCD phase diagram as a function
of the chemical potential $\mu$. From $\mu=0$ to the vertical dashed line at
$\mu \approx 0.22$ GeV, the transition is a cross-over. The height of the
heat capacity, the thermodynamic variable drawn on the vertical line, becomes
singular as it approaches the dashed line, suggesting a change in the nature of
phase transition from a simple cross-over to the first order phase transition.
The chiral symmetry continues to be broken above $\mu \approx 0.22$ GeV, as
evidenced from Fig.~\ref{fig3}. However, it becomes numerically demanding to
find the transition points. Chiral symmetry appears to be restored at
$\mu \approx 0.53$ GeV, represented by a vertical dot-dashed line.
\label{fig6}}
\end{figure}
Notice from Fig.~\ref{fig3} that chiral symmetry remains broken beyond $\mu \approx 0.22 GeV$, where the first order phase
transition is triggered. Our numerical accuracy allows us to explore the chemical
potential up to about 0.3 GeV. However, the trajectory of the points
along the critical line is smooth enough to make a quadratic fit to estimate the value $\mu \approx
0.53\,\rm GeV$ for zero temperature where chiral symmetry is finally restored, as shown in Fig.~\ref{fig5}. All these characteristics are sketched in Fig.~\ref{fig6}.

A few words of caution are here in order. The function $D(T)$ in Eq.~(\ref{V-ansatz})
and the full quark-gluon vertex $\Gamma_\nu(\vec{q},\tilde{w}_l,\vec{p},\tilde{w}_n)$
surely depend on $\mu$ and a different choice of these quantities may change the
position of the critical end point. Moreover, the fermion-boson vertex
must be constructed from theoretical~\cite{T-vertex}, as well as phenomenological~\cite{P-vertex}
constraints. At finite $T$ and (or) $\mu$, the complexities involved
are highly non-trivial but work has already begun in this direction~\cite{Ayala:2003}.
We have performed preliminary studies to explore the nature of the confinement-deconfinement transition
and have found that  within this model the transition line follows the footsteps of the chiral
symmetry breaking-restoration transition. This indicates that other transitions to quarkionic
matter or a color locked phase, as the chemical potential increases, can only be described by
incorporating additional ingredients to this model.
While the above mentioned details will certainly be trimmed over the next few years,
this letter shows that SDEs are an efficacious tool to explore the QCD
phase diagram.  We have used this tool to locate the critical end point as well as to provide clear signals
of the cross over (near $\mu=0$), and of a first order phase transition as we move away to sufficiently
high values of $\mu$, starting from lattice results for the finite temperature quark-antiquak condensate.

\noindent
{\bf Acknowledgments:} This work has been supported by
CIC-UMICH grants 4.10 and 4.22, CONACyT
grant numbers 82230 and 128534 and DGAPA-UNAM grant number IN103811.

\end{document}